\documentclass[12pt, letterpaper]{article}
\usepackage[margin=2cm]{geometry} %for arxiv

\usepackage{amssymb}
\usepackage{amsmath}
\usepackage{graphicx}
\usepackage{physics}

\usepackage{todonotes}

\newcommand{\Gr}{\mathsf{G}}
\newcommand{\pp}{\mathsf{p}}

\usepackage{cite}
\usepackage{url}
\newcommand{\subs}[1]{\textbf{#1}}

\usepackage{color}
\newcommand{\change}[1]{#1}

\title{Entropic dynamics of networks}

\author{Felipe Xavier Costa$^1$, Pedro Pessoa$^1$ \\
$^1$Department of Physics, University at Albany - SUNY \\  Albany, NY - USA}
\date{}

\begin{document}
\maketitle
\abstract{
Here we present the entropic dynamics formalism for networks.
That is, a framework for the dynamics of graphs meant to represent a network derived from the principle of maximum entropy and the rate of transition is obtained taking into account the natural information geometry of probability distributions.
We apply this framework to the Gibbs distribution of random graphs obtained with constraints on the node connectivity. The information geometry for this graph ensemble is calculated and the dynamical process is obtained as a diffusion equation.
We compare the steady state of this dynamics to  degree distributions found on real-world networks. 
}


\newpage
%\tableofcontents
\section{Introduction} \paragraph{}

 Since the work of Jaynes \cite{Jaynes57,Jaynes57b}, the original method of maximum entropy (MaxEnt) has explained thermodynamics in terms of information theory by deriving Gibbs distributions in the context of statistical mechanics. Those distributions arise as a result of a well-posed problem, namely selecting the distribution that is least-informative under a set of expected value constraints. Also, the Gibbs distributions coincide with what is known in statistical theory as the exponential family -- the only distributions for which a finite set of sufficient statistics, functions that generate the expected values, exist (see e.g. \cite{DKPTheorem}). 
Under this general understanding, it is not surprising that, after Jaynes, MaxEnt has been presented as a general method for inference  \cite{ShoreJohnson80,Skilling88,Caticha03,Vanslette17} and being applied in a large range of subjects such as, but not limited to,  economics \cite{Golan08,CatichaGolan14}, ecology \cite{Harte11,Bertram19}, cell biology \cite{Martino18,Dixit20}, opinion dynamics \cite{Vicente14,Alves16} and geography \cite{Wilson67,Yong16}. 
This perspective of MaxEnt is seen as a method for updating probability distributions when new information about the system becomes available. Under this understanding it is also not surprising that the methods of Bayesian inference \cite{Giffin06} and  several machine learning techniques \cite{Barber12}, including those used in image processing \cite{Skilling84,Higson18} and deep learning \cite{Goodfellow16,Bahri20} are found to be a particular application of MaxEnt -- either as a consequence of Bayes or directly.

Gibbs distributions have been studied in the context of random graphs by Park and Newman \cite{Park04}, leading to an extensive investigation of MaxEnt applications in network science \cite{Bianconi07,Bianconi09,Anand09,Anand11,Peixoto12,Cimini19,Radicchi20}.
In this plethora of investigations, many models are proposed as different choices of the sample space -- e.g. simple graphs, weighted graphs -- and sufficient statistics -- functions defined over the possible graphs  (e.g. total connectivity, node degrees sequences, and average nearest neighbour connectivity). 
However, considering these MaxEnt procedures, when one chooses a function as sufficient statistics it does not mean that a precise number is known for their expected values.
As an example, MaxEnt obtains the correct distribution of link placement for these scale-free networks when one chooses the degree of each node as sufficient statistics, the power law behavior for the node degrees is fitted posteriorly with data.
However, as explained by Radicchi \emph{et al.} \cite{Radicchi20}, this MaxEnt model alone does not justify why in some networks the node degrees are highly heterogeneous. Therefore an external model for sampling degree sequences is needed.

The issue described is a statement of the fact that, despite its generality, MaxEnt cannot tell by itself what constraints are relevant to the specific problem. The accuracy of constraints are justified by the fact that they work in practice, that is, they lead to a model that accurately describes the system of interest. For example, in physics \cite{Jaynes65} we can assume that the microscopic world follows a conservative (Hamiltonian) dynamic, leading to expected value constraint of conserved quantities. Ultimately, unavoidable scientific labor is necessary to understand which constraint correct implement the information one has about the system of interest.

On the other hand, the principles of information theory can be used to obtain the laws of dynamics for stochastic dynamical systems --  the transition probabilities are derived by maximizing an entropy -- this idea is referred to as entropic dynamics (EntDyn) \cite{Caticha10} and has already found successful applications in quantum mechanics\footnote{For the applications of EntDyn in quantum mechanics and quantum field theory the constraints must be chosen so that the Hamiltonian structure is recovered.} \cite{Caticha19}, quantum fields \cite{Ipek19}, renormalization groups \cite{Pessoa18}, finance \cite{Abedi19a}, and neural networks \cite{NCaticha20}. 
In a recent paper \cite{Pessoa20} we presented an entropic formalism for dynamics in a space of Gibbs distributions. The dynamics developed there relies on the concepts of information geometry \cite{Caticha15,Amari16,Ay17,Nielsen20}, an area of investigation that assigns differential geometric structure to the space of probability distributions\footnote{Incidentally, information geometry has been applied to generate measures for complexity, see e.g. \cite{Ay11,Felice14,Franzosi16,Felice18}.}.
The dynamics obtained is a diffusion process in the  Gibbs statistical manifold - space of Gibbs distributions parametrized by the expected values and endowed with a Riemmanian metric from information geometry.

A widely known example for dynamics of networks considers a preferential attachment mechanism for the evolution of node degrees(see e.g \cite{Barabasi99,Bianconi01,Albert02}), leading to scale-free networks, where the degree distribution follows a power law.
On the other hand, there has been
literature challenging reported scale-free networks \cite{Broido19} and power laws in general \cite{Clauset09}. These argue that one can not verify the power law behavior in real world networks under strong statistical validation, which indicates that scale-free networks are expected from a highly idealized processes and further dynamical models accounting for the peculiarities of a particular system are in order \cite{Barabasi18,Holme19}.

Our goal with the present article is to show how EntDyn can provide a systematic way to derive dynamics of network ensembles. As an example, we apply the EntDyn developed in \cite{Pessoa20} to the space of Gibbs distributions of graphs obtained after choosing the node degrees as constrains. 
We compare the steady state distributions obtained from EntDyn to the distributions found in real-world networks \cite{Broido19}. These results are not based on an underlying dynamics with particular assumptions, rather they are a consequence of the information geometry of networks ensembles.
{We bring into attention} that although the dynamics developed here is simple, we also comment on how the framework provided by EntDyn is flexible enough so that further constraints can be implements to account for the information available about the dynamical process.

In the following section, we will present the random graphs model used and the maximum entropy distributions and information geometry derived from it. 
In section 3, we review the entropic dynamics presented on \cite{Pessoa20} in the context  of random graphs obtaining a differential equation for the dynamics of networks.
In section 4, we find the steady-states of the differential equation and argue on how the power law behavior emerges from the dynamics.

\section{Gibbs distributions of graphs} \paragraph{}
In this section, we will establish  the random graph model for the present article and obtain the Gibbs distribution and the metric tensor for its information geometric structure. A graph is defined by a set of nodes (or vertices) $V$ and a set of links (or edges) $E  \change{=\{\varepsilon_\mu\} }$. Each link connects two nodes, \change{thus, we will write} $\varepsilon_\mu=({i,j})$, where $i,j \in \{1,2,\ldots,|V|\}$ are elements of an enumeration for the set of nodes $V.$ 
In network science, meaning is attached to the elements of a graph models as its nodes represent entities and its links represent interactions between entities (for example networks where the links represent publication coauthorships between scientists -- nodes -- \cite{Molontay19},  links representing associations between gene/disease \cite{Goh07} or scientific concepts \cite{Stella19}). For the scope of the present article we treat graphs in a general manner without attributing any information related to what the random graph may represent.
Because of this, the constraint defined here and the dynamical assumptions in the next section will be as general as possible.

\subsection{MaxEnt of graphs} \paragraph{}
At this stage we will attribute, through MaxEnt, a probability distribution $\rho(\Gr)$ for each graph (microstate) $\Gr = (V,E)$. Inspired by Radicchi \emph{et al.} \cite{Radicchi20} -- although similar descriptions have been proposed before e.g. \cite{Park04,Bianconi09} -- we will suppose a graph with $N = |V|$ nodes and $L =|E|$ \change{links}, and constraints on the number of \change{links} (also referred to as degree or connectivity) of each node $i$.

To obtain the appropriate distribution $\rho(\Gr)$  we ought to maximize the functional

\begin{equation}\label{KLentropy}
    S[\rho|q] = - \sum_\Gr \rho(\Gr) \log \frac{\rho(\Gr)}{q(\Gr)} = - \sum_{E} \rho(E) \log {\rho(E)}  \ ,
\end{equation}
where $q$ is a prior distribution. The functional $S[\rho|q]$ in \eqref{KLentropy} is known as Kullback-Leiber (KL)
\footnote{Even though the network problem might make one expect to obtain power laws it does not mean we should skew away from KL entropy. It has been widely reported that other functionals proposed to replace KL entropy, such as Renyi's or Tsallis', induce correlations not existing in the prior or constraints \cite{Presse13,Presse14,Oikonomou19,Pessoa20b} and therefore lead to inconsistent statistics. } 
entropy, reducing to Shannon entropy when $q$ is uniform.
The last equality in \eqref{KLentropy} holds as we assume that we are inferring over an already known number of nodes $N$ and  uniform prior.
Each link is treated independently under the same constraints of node connectivity. Since Shannon entropy is additive -- meaning that for independent subsystems the joint entropy is the sum of entropies calculated for each subsystem -- and preserves subsystems independence \cite{Vanslette17} -- meaning if two subsystems are independent in the prior and the constrains do not require correlations between the posterior (distribution that maximizes entropy) will also be independent for each subsystem -- therefore, with separate constraints for each link, \eqref{KLentropy} reduces to
\begin{equation} \label{intermediate}
    S[\rho|q] =  -\sum_E \left( \prod_\mu \rho(\varepsilon_\mu) \right) \log \left( \prod_\mu \rho(\varepsilon_\mu) \right)  \change{ = - \sum_\mu \sum_{\varepsilon_\mu} \rho(\varepsilon_\mu) \log(\rho(\varepsilon_\mu)) } \ . %L  \ s[\rho]\ ,
\end{equation}
\change{
If the same constraint is applied to each link, MaxEnt results in distributions of the same form for each $\varepsilon_\mu$. Because of this, we can treat each link separately and write, for simplicity, $\rho(\varepsilon) = \rho(\varepsilon_\mu)$ for every $\mu$. So that \eqref{intermediate} can be written as
}
\begin{equation} \label{Spn}
    S[\rho|q] = L \ s[\rho] \ , \qq{where}  s[\rho] \doteq -  \sum_\varepsilon  \rho(\varepsilon) \log\rho(\varepsilon) \ ,
\end{equation}
 \change{ and $s$ can be referred to as the entropy per link.}
Thus maximizing entropy  $S$ for the graph $\Gr$ is equivalent to maximizing $s$ \change{for a link} $\varepsilon$.

To implement, in our MaxEnt procedure, that the relevant information  is the degree of each node we \change{define the functions}  $ a^i(\varepsilon = (j,m)) \change{\doteq} \frac{1}{2}({\delta^i_j + \delta^i_m})$ \change{-- where $\delta^i_j$ refers to the Kronecker delta\footnote{\change{That means, $\delta^i_i = 1$ and $\delta^i_j = 0$ if $i\neq j$. Equivalently $\delta_{ii} = 1$ and $\delta_{ij} = 0$ if $i\neq j$.  }} 
--   as our}  sufficient statistics. Leading to the expected value constraints 
\begin{equation} \label{MEconstraint}
    \change{\expval{a^i(\varepsilon)}} = \sum_{j,m} \rho(\varepsilon=(j,m)) \ \left( \frac{\delta^i_j + \delta^i_m}{2} \right) = \frac{k^i}{2L} = A^i \ ,
\end{equation}
where $k^i$ is the expected degree of each node $i$. The $2L$ factor is included so that the expected values $A^i$ sum to unity, since by construction, the sum of degrees is twice the number of \change{links}.
The function that maximizes \eqref{Spn} under \eqref{MEconstraint} and normalization  is the Gibbs distribution
\begin{equation}\label{distribution}
\rho(i, j | \lambda) = \change{\frac1Z \exp\left(-\sum_m \lambda_m a^m(i,j)\right) = } \frac1Z e^{-\frac12 \lambda_i - \frac12\lambda_j} \ , 
\end{equation}
\change{where $Z$ is the normalization factor}
\begin{equation}
Z = \sum_{i, j} e^{-\frac12\lambda_i - \frac12\lambda_j} = \qty(\sum_i  e^{-\frac12\lambda_i})^2 \ ,
\end{equation}
and $\lambda = \{\lambda_1, \lambda_2, \ldots, \lambda_N\}$ is the set of Lagrange multipliers dual to  the expected values $A = \{A^1, A^2,  \ldots, A^N\}$. 
In \eqref{distribution}, we have the probability that a link $\varepsilon$ connects the nodes $i, j$ given the set of Lagrange multipliers. However, \eqref{MEconstraint} indicates that they can also be parametrized by the expected values $A$. The two sets of parameters are related by
%In \eqref{distribution} we use $\rho(i,j)$ as the probability that a link $\varepsilon$ connects the nodes $i$ and $j$, $\rho(\varepsilon = (i,j))$. Per \eqref{distribution} each set $\lambda$ leads to an unique Gibbs distribution, at the same time \eqref{MEconstraint} indicates that they can also be parametrized by the expected values $A^i$. The two sets of parameters are related by
\begin{equation}
    A^i = - \frac{1}{Z} \pdv{Z}{\lambda_i} = \frac{ e^{-\frac12\lambda_i}}{\sum_j  e^{-\frac12\lambda_j}} \ .
\end{equation}
Equating the previous result with  \eqref{MEconstraint} we obtain $k^i = e^{-\frac12\lambda_i}$ allowing us to write
\begin{equation}\label{AiAj}
    \rho(\varepsilon = (i,j)|A) = \frac{k^i k^j}{(2L)^2} = A^iA^j \ .
\end{equation}
That is, we can interpret $A^i$ as the probability for which a specific link $\varepsilon$ has the node $i$ in one of its ends. 
For reasons that will be presented later in our investigation, it is also useful to calculate the node entropy at its maximum as a function, rather than a functional, of the expected values, meaning
\begin{equation}\label{graphentropy}
    s(A) \doteq s[\rho(i,j|A)] = - 2 \sum_i A^i \log A^i \ ,
\end{equation}
the last equality is found by substituting \eqref{AiAj} into \eqref{Spn}. %We will refer to $s(A)$ as the graph entropy.

Since we can parametrize the space of probability distributions by expected values $A^i$ we will use those as coordinates when assigning the geometry to this space in the following subsection. 

\subsection{Information Geometry}
\paragraph{}
Our present goal is to assign a Riemmanian geometric structure 
\change{to probability distributions, this is achieved by Information Geometry. For the scope of the present article only the geometric structure of}
the distributions defined in \eqref{AiAj} \change{will be studied\footnote{\change{For more general and pedagogical texts on information geometry see e.g. \cite{Caticha15,Amari16,Ay17,Nielsen20}}}. From \eqref{AiAj}} the space of Gibbs distributions is \change{parametrized} by the values of $A$, and the distances obtained from 
$\dd \ell^2 = \sum_{i,j} g_{ij} \dd A^i \dd A^j$ are a measure of distinguishability between the neighbouring distributions $\rho(i,j|A)$ and $\rho(i,j|A+\dd A)$. The metric components $g_{ij}$ are given by the Fisher-Rao information metric (FRIM) \cite{Fisher25,Rao45}

\begin{equation}
\label{frim}
    g_{ij} = \sum_{m,n} \ \rho({m,n}|A) \frac{\partial\log \rho({m,n}|A)}{\partial A^i}\frac{\partial\log \rho({m,n}|A)}{\partial A^j}  .
\end{equation}
This metric is not arbitrarily chosen, FRIM provides the only Riemmanian geometric structure that is consistent with Markov embeddings \cite{Cencov81,Campbell86},   hence this metric structure is a consequence of the grouping property of probability distributions.

Before calculating the FRIM for the distributions defined in \eqref{AiAj} \change{it is} important to remember that $\sum_i A^i = 1$. We will express then the value related to the last node in the enumeration as a dependent variable $ A^N= 1-\sum_{i=1}^{N-1} A^i$. Since we are describing graphs with a fixed number of links $L$, the constraints defined in \eqref{MEconstraint} have some level of redundancy, namely $A^N$ is automatically defined by the set of all others. Even though this does not interfere with the maximization process -- MaxEnt is robust enough  to properly deal with redundant information -- this has to be taken into account when calculating the summations in \eqref{frim}.

Therefore  the FRIM components for the probabilities obtained in \eqref{AiAj} are
\begin{equation}
    g_{ij} = \frac{2\delta_{ij}}{A^i} + \frac{2}{A^N} \quad \text{where} \quad  i,j \in [1,N-1] \ .
\end{equation}
It would be useful to have an expression valid for all indexes, $i,j \in [1,N]$. For it we use, as in \cite{Caticha15}, that  $ \dd A^N= -\sum_{i=1}^{N-1} \dd A^i$, the expression for infinitesimal distances then becomes
\begin{equation}
    \dd \ell^2 = \sum_{i=1}^{N-1} \sum_{j=1}^{N-1} \left( \frac{2\delta_{ij}}{A^i} + \frac{2}{A^N} \right) \dd A^i \dd A^j  = \sum_{i,j} \frac{2\delta_{ij}}{A^i} \dd A^i \dd A^j \ .
\end{equation}
Yielding then a much simpler and diagonal metric tensor 

\begin{equation} \label{finalmetric}
    g_{ij} = \frac{2}{A^i}\delta_{ij} \ .
\end{equation}
As it is a property of Gibbs distributions \cite{Caticha15,Pessoa20} this metric tensor could also have been found as the Hessian of $- s(A)$ in \eqref{graphentropy}. The diagonal \change{metric obtained} is consistent with the fact that, per \eqref{AiAj}, both nodes at the end of a link will be sampled independently with the same distribution.

Having calculated the metric for the Gibbs distributions of our graph model, we have all elements to define a dynamics on it in the following section.

\section{Entropic dynamics of Gibbs distributions} \paragraph{}

Entropic dynamics is a formalism for which the laws of dynamics are derived from entropic methods of inference. For the scope of the present article we are going to evolve  the parameters $A$ representing a change for the probabilities for $\varepsilon$ in \eqref{AiAj}, this is equivalent to  have distributions from which the sequences of node degrees, $k^i = 2L A^i$, are sampled. In this description, the probabilities of links can be recovered from

\begin{equation} \label{jointprobability}
    P(\varepsilon,A) =   P(A) \rho(\change{\varepsilon}|A) \ ,
\end{equation}
where $\rho$ is defined in \eqref{AiAj}.
The dynamical process will describe how the change from a set of parameters $A$ -- representing an instant of the system-- evolves to a set of parameters $A'$ for which  the distribution for a later instant $P(A')$ is assigned as

\begin{equation}
    P(A') = \int \dd A \ P(A'|A) P(A)
\end{equation}
EntDyn consists on finding the transition probability $P(A'|A)$ through the methods of information theory. As done in our previous work \cite{Pessoa20}, the dynamical process will rely on two assumptions: 
(i) the changes happens continuously\footnote{Continuous motion might not sound as a natural assumption in a discrete system, such as graphs, however even if a space is discrete, the set of probability distributions on it is continuous as are the expected values that parametrize it.} 
which will determine the choice of prior and (ii) that the motion has to be restricted to the Gibbs distributions obtained from $\rho(\varepsilon|A)$ in \eqref{AiAj} which will determine our constraint.
Beyond the scope of the present article, different models can be generated by imposing constraints that implements other information known about the dynamical process.

\subs{The entropy} we need to maximize has to account for the joint change in the degrees of uncertainty in the the graph \change{links} $\varepsilon$ as well as  the parameters $A$ which is represented by the distribution in \eqref{jointprobability}.  
The transition from $A$ to $A'$ must also contain information about the transitions from $\varepsilon$ to a latter link distributions $\varepsilon'$. Therefore, we must maximize entropy for the joint transition $P(\varepsilon', A' \mid \varepsilon, A)$, meaning

\begin{equation}
\mathcal{S}[P| Q] = - \sum_{\varepsilon'}\, \int \dd A'\, P(\varepsilon', A' \mid \varepsilon, A) \log(\frac{P(\varepsilon', A' \mid \varepsilon, A)}{Q(\varepsilon', A' \mid \varepsilon, A)}) \ .
\label{DynamicalS}
\end{equation}
where $Q(\varepsilon', A' \mid \varepsilon, A)$ is the prior to be determined. 
We shall call $\mathcal{S}$ the dynamical entropy to avoid confusion with the graph entropy in \eqref{KLentropy} and the entropy per link \eqref{Spn}.

\subs{The prior} 
that implements continuity for the motion on the statistical manifold but is otherwise uninformative is of the form
\begin{equation}
Q(\varepsilon', A' \mid \varepsilon, A) = Q(\varepsilon' \mid \varepsilon, A, A') Q(A' | \varepsilon, A) \propto  g^{1/2}(A') \exp(-\frac1{2\tau} \change{\sum_{ij}} \ g_{ij} \Delta A^i \Delta A^j) \ ,
\label{eq:EDGPrior}
\end{equation}
as explained in \cite{Pessoa20}, where $\Delta A^i = {A'}^i-A^i$, $g=\det g_{ij}$, and ${\tau}$ is a parameter that will eventually take the role of time,
since when $\tau \rightarrow 0$ leads to short steps, meaning
$d\ell \rightarrow 0$.

\subs{The constraint} that implements that the motion does not leave the space of Gibbs distributions defined in Section 2.1 is

\begin{equation}
P(\varepsilon', A' \mid \varepsilon, A) = P(\varepsilon' \mid \varepsilon, A, A') P(A' | \varepsilon, A) = \rho(\varepsilon' \mid A') P(A' | \varepsilon, A) \ . 
\label{eq:EDGConst}
\end{equation}
that means the distribution for $\varepsilon'$ conditioned on $A'$ must be of the form \eqref{AiAj}.
Note that per \eqref{eq:EDGConst} 
the only factor still undetermined for the full transition probability is $P(A'| \varepsilon, A)$.

\subs{The result} obtained when maximizing \eqref{DynamicalS} with the prior \eqref{eq:EDGPrior} and under \eqref{eq:EDGConst} is

\begin{equation} \label{result1}
    P(A' | \varepsilon, A) \propto  g^{1/2}(A') \exp(s(A') -\frac1{2\tau} \change{\sum_{ij}} \ g_{ij} \Delta A^i \Delta A^j) \ .
\end{equation}
Note that it is independent of $\varepsilon$, which is not surprising since neither the prior nor the constraints assume any correlation between $A'$ and $\varepsilon$ \change{thus}, by marginalization,  $P(A' | A)=P(A' | \varepsilon, A)$.
For short steps, $\dd \ell \rightarrow 0$ therefore $\Delta A \rightarrow 0$, we can expand $s$ in the linear regime, leading to the transition probability of the form 

\begin{equation}
P(A' | A) = \frac{1}{\mathcal{Z}(A)} g^{1/2}(A') \exp(\sum_{i}\pdv{s}{A^i} \Delta A^i -\frac1{2\tau} \sum_{ij} g_{ij} \Delta A^i \Delta A^j) \ ,
\end{equation}
where the normalization factor $\mathcal{Z}(A)$ absorbs the proportionality constant in \eqref{result1} and $e^{s(A)}$. 
In \cite{Pessoa20} we calculate the moments for this transitions up to order $\tau$ obtaining
\begin{equation} \label{smooth}
\begin{split}
        \expval{\Delta A^i} = \tau \sum_{j,k}\left( g^{ij}\pdv{s}{A^j} - \frac{1}2\Gamma^i_{jk}g^{jk} \right) \ ,  \\
        \expval{\Delta A^i\Delta A^j} = \tau g^{ij} \ ,  \quad \text{and} \quad \expval{\Delta A^i \Delta A^j \Delta A^k} = 0 \ ;  
\end{split}
\end{equation}
where $g^{ij}$ are the elements of the inverse matrix to $g_{ij}$, $\sum_j g_{ij} g^{jk}  = \delta_i^k$ and $\Gamma^i_{jk}$ are the Christoffel symbols.%\todoblue{forgot the detail, should we talk more about CS?}

Equation \eqref{smooth} is the definition of a smooth diffusion \cite{Nelson85} if we choose $\tau$ as a time duration $\Delta t$, which is equivalent to calibrating our time parameter in terms of the fluctuations $\Delta t \doteq \tau \propto \sum_{ij} g_{ij} \Delta A^i \Delta A^j$. That means, here the role of time  emerges from  emergent properties of the motion, up to a multiplying constant, time measure the fluctuations in $A$. The system is its own clock. 
As explained in \cite{Pessoa20}  this leads to the evolution as a Fokker-Planck equation 

\begin{equation}\label{FPeq}
\pdv{p}{t} = - \frac1{g^{1/2}} \sum_i  \pdv{A^i}\qty(g^{1/2} p  v^i)\ , \qq{where} v^i = \change{\sum_{j}\ } g^{ij} \pdv{A^j}\qty(s - \frac12\log(p)) \ ,
\end{equation}
and $p$ is the invariant probability density $p(A) \doteq \frac{P(A)}{\sqrt{g(A)}}$. If we substitute the graph entropy in \eqref{graphentropy} and the metric in \eqref{finalmetric} we obtain

\begin{equation} \label{diferentialevolution}
\pdv{p}{t} = \sum_i \left( \qty[\frac12\log{A^i} + \frac3{2}]p + \qty[A^i\qty(\log{A^i} + 1) + \frac1{8} ]\pdv{p}{{A^i}} + \frac{{A^i}}4\pdv[2]{p}{{A^i}} \right)  \ .  
\end{equation}
This establishes the dynamical equation for the graph model. In the following section we will focus on finding a steady-state $\Bar{p}(A)$ for \eqref{diferentialevolution}.

\section{Entropic dynamics of graphs models}
\paragraph{}
In order to find a steady state in \eqref{diferentialevolution} is interesting to see that for $\pdv{\Bar{p}}{t} = 0$ the equation is separable, meaning the solution can be written as a product of the same function, $\pp(a)$, for each argument $\Bar{p}(A=\{A^i\}) = \prod_i \pp(a=A^i)$. This leads to say that each term in the summation on \eqref{diferentialevolution} has to be zero and therefore $\pp(a)$ has to follow 

\begin{equation} \label{separatedfora}
 \frac{a}4 \dv[2]{\pp}{{a}} +   \qty[a\qty(\log{a} + 1) + \frac1{8} ]\dv{\pp}{{a}} + \qty[\frac12\log{a}+ \frac3{2}]\pp  = 0  \ .  
\end{equation}
In order to solve the above equation we make the substitution $y=\sqrt{8a}$, so that it transforms into

\begin{equation} \label{separatedfory}
 \dv[2]{\pp}{{y}} +   y\qty[f(y) -2] \dv{\pp}{y} +f(y) \pp = 0 \ ,
\end{equation}
where $f(y) = 2 \log y - \log 8 +3$.
This substitution is equivalent to write \eqref{FPeq} under a change of coordinates $Y^i = \sqrt{8A^i}$, in which the metric \eqref{finalmetric} transforms into an Euclidean metric, $\sum_{ij} g_{ij} \dd A^i \dd A^j = \sum_{ij} \delta_{ij} \dd Y^i \dd Y^j$.

The range at which \eqref{separatedfory} is valid takes into account the fact that the maximum possible connectivity is the number of links, $k=L$, leading to a possible maximum value for $a = 1$ and $y=\sqrt{8}$,  when self-connections are not ignored. However, Anand \textit{et al.} \cite{Anand11} argues that in order for the connectivity of each node to remain uncorrelated, a lower maximum connectivity should be considered. Inspired by their arguments we can set $y_\text{max}=\sqrt[4]{2^5/L}$, corresponding to $k_{max}= \sqrt{2L}$ and $a_{max} = 1/\sqrt{2L}$. Also, we can see that \eqref{separatedfory} diverges at $y=0$ unless $\pp(y=0) = 0$. Therefore, we'll consider $y \in (0, y_{\text{max}}]$.

Solving \eqref{separatedfory} is enough to obtain the steady-state values for $\pp(a)$ in \eqref{diferentialevolution} and therefore the degree distribution
\begin{equation}\label{DegDistDef}
P\left(a = \frac{k}{2L}\right) = \sqrt{\frac2{a}} \pp\left(y = \sqrt{8a}\right) \ ,
\end{equation}
where the square root factor comes from the information metric $\sqrt{g(a)} = \sqrt{\frac2{a}}$. 
We choose to solve \eqref{separatedfory} as an initial value problem (IVP) by making sure that the node connectivity remains uncorelated, thus setting
\begin{equation}\label{IVPYM}
\pp(y=y_{\text{max}}) = 0 \qq{and} \eval{\dv{\pp}{y}}_{y=y_{\text{max}}} = \dd p_0 \ ,
\end{equation}
where $y_\text{max}=\sqrt[4]{2^5/L}$ was considered for $L = 2^9\text{, } 2^{13}\text{ and } 2^{30}$. The final result is later normalized, and the precise value of $\dd p_0$ is to be investigated.

\begin{figure}%[h]
\centering
  \begin{minipage}[b]{0.31\textwidth}
    \includegraphics[width=\textwidth]{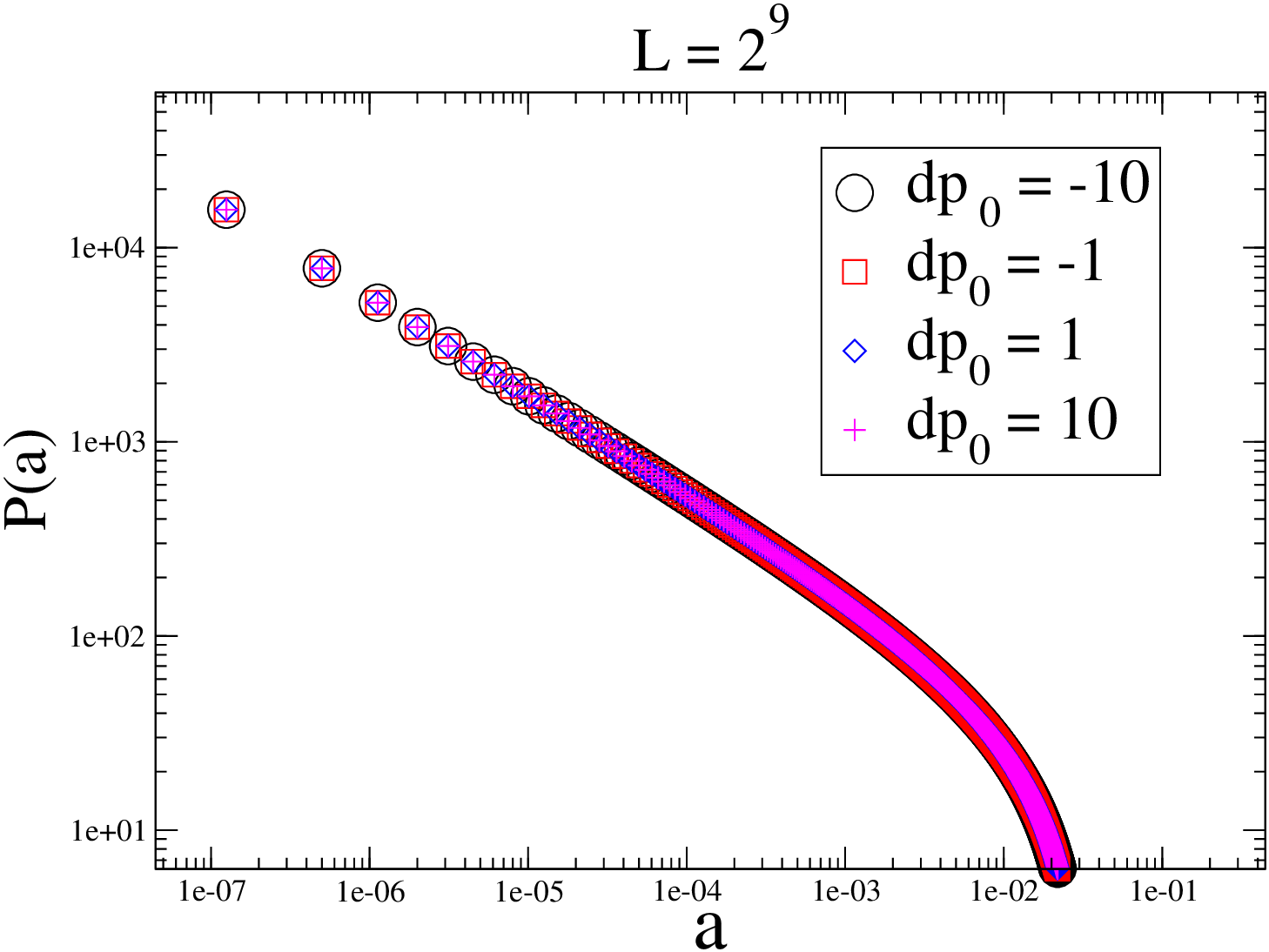}
  \end{minipage}
  \begin{minipage}[b]{0.31\textwidth}
    \includegraphics[width=\textwidth]{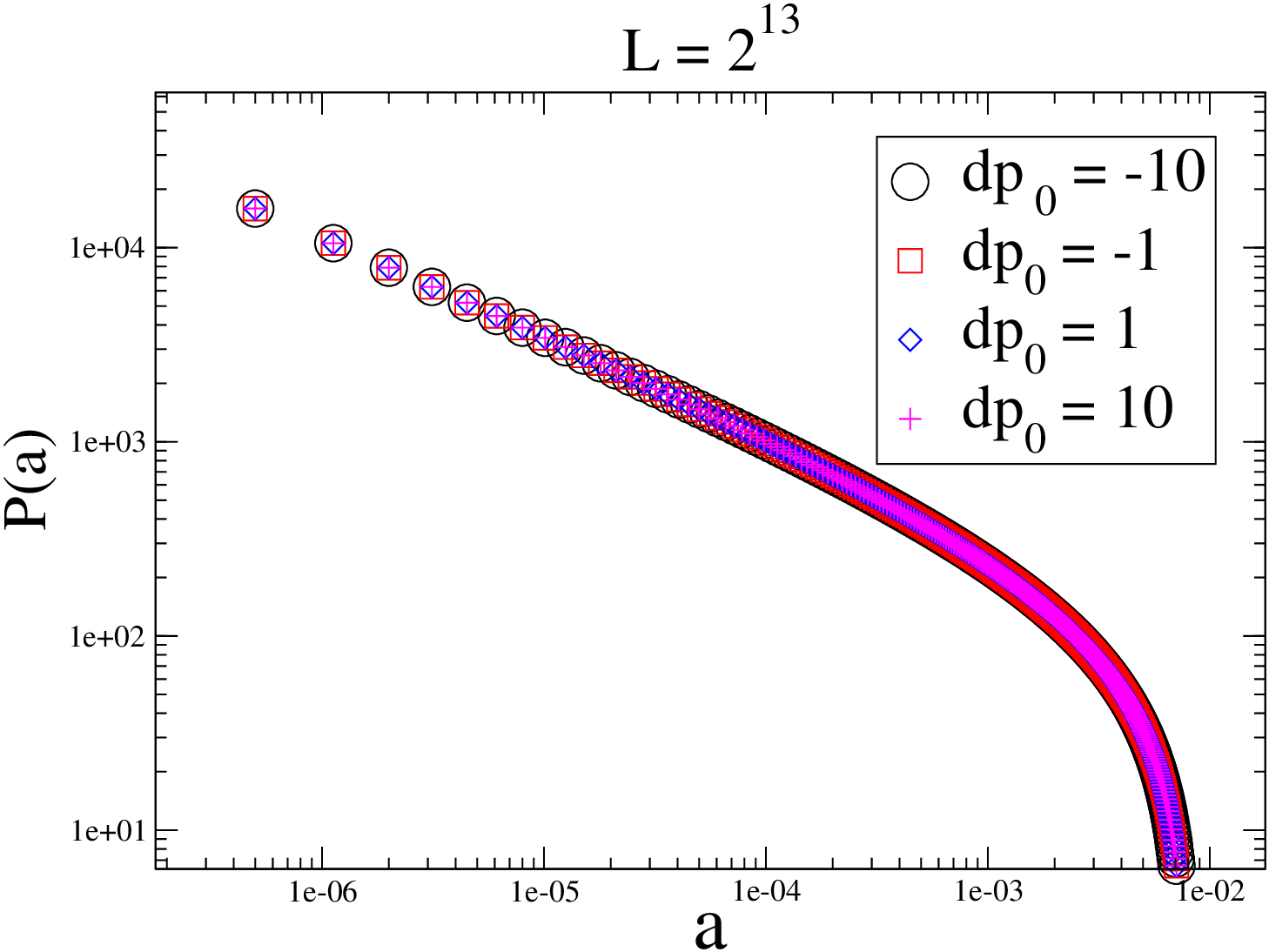}
  \end{minipage}
   \begin{minipage}[b]{0.31\textwidth}
    \includegraphics[width=\textwidth]{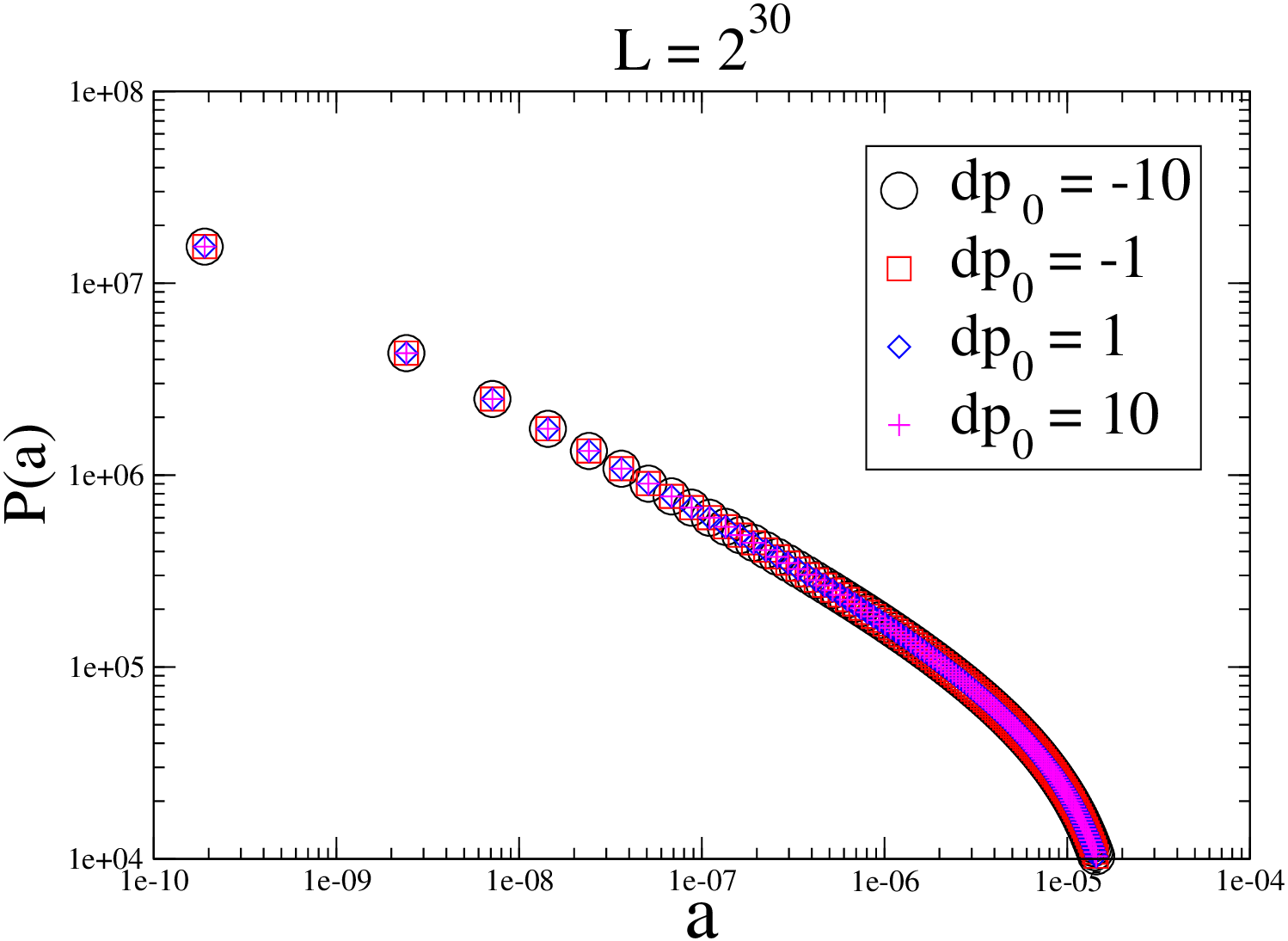}
  \end{minipage}
  \caption{Degree distribution derived from the solution of the IVP \eqref{IVPYM} for different values of $L = 2^9$ (left), $2^{13}$ (center), and $2^{30}$ (right). The degree distribution does not depend on $\dd p_0$ and have a very similar behaviour across $L$.}
  \label{fig:IVPLYMax}
\end{figure}

\begin{figure}%[h]
  \vspace{.5cm}
  \begin{minipage}[b]{0.45\textwidth}
    \includegraphics[width=\textwidth]{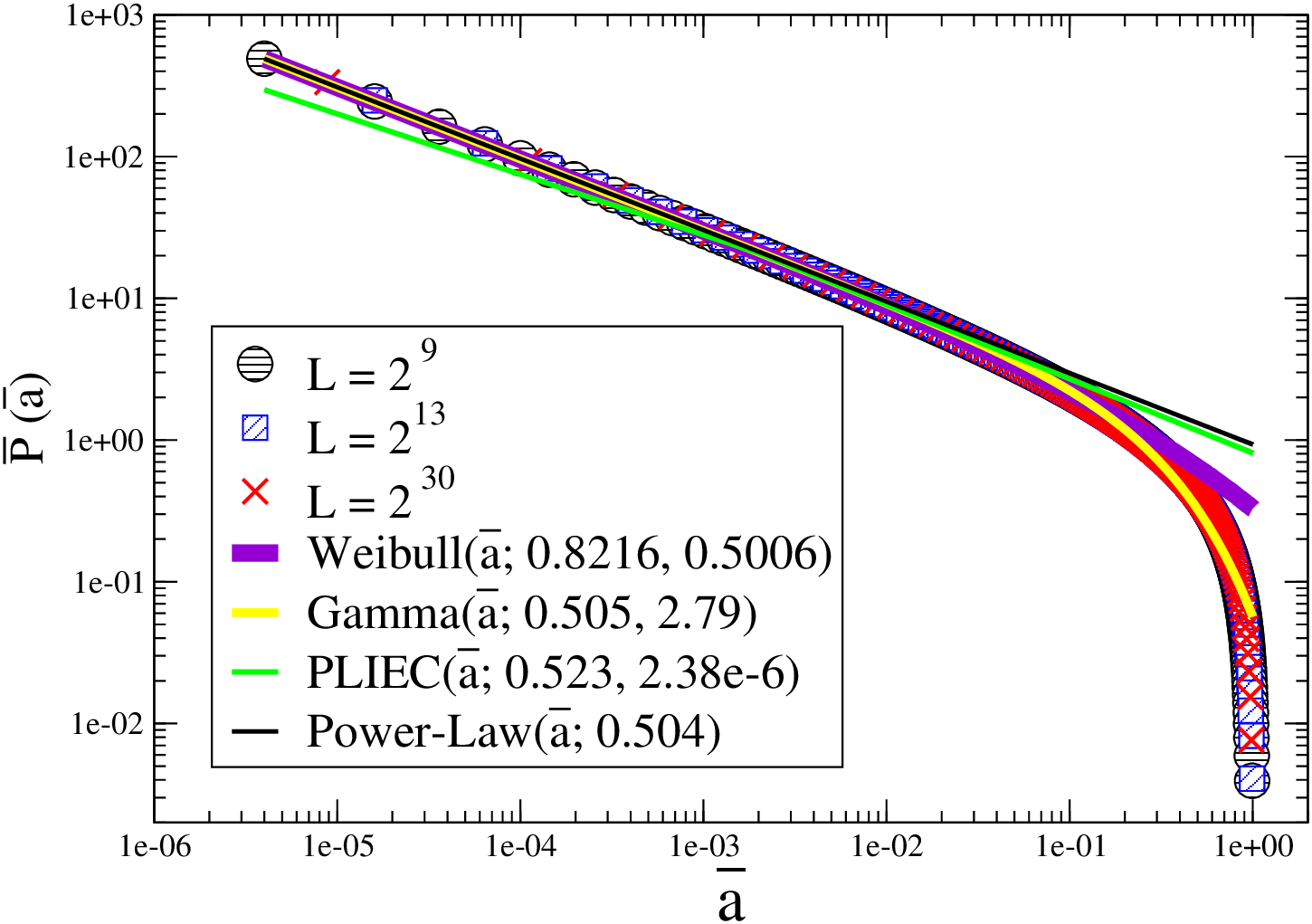}
    \caption{Re-scaled degree distribution for the IVP \eqref{IVPYM} irrespective of the number of links $L$. The result fits well with a Weibull and Gamma, also known as power law with cutoff, distributions within most of the allowed range.}
     \label{fig:IVPL32} 
  \end{minipage}
  \hfill
  \begin{minipage}[b]{0.45\textwidth}
    \includegraphics[width=\textwidth]{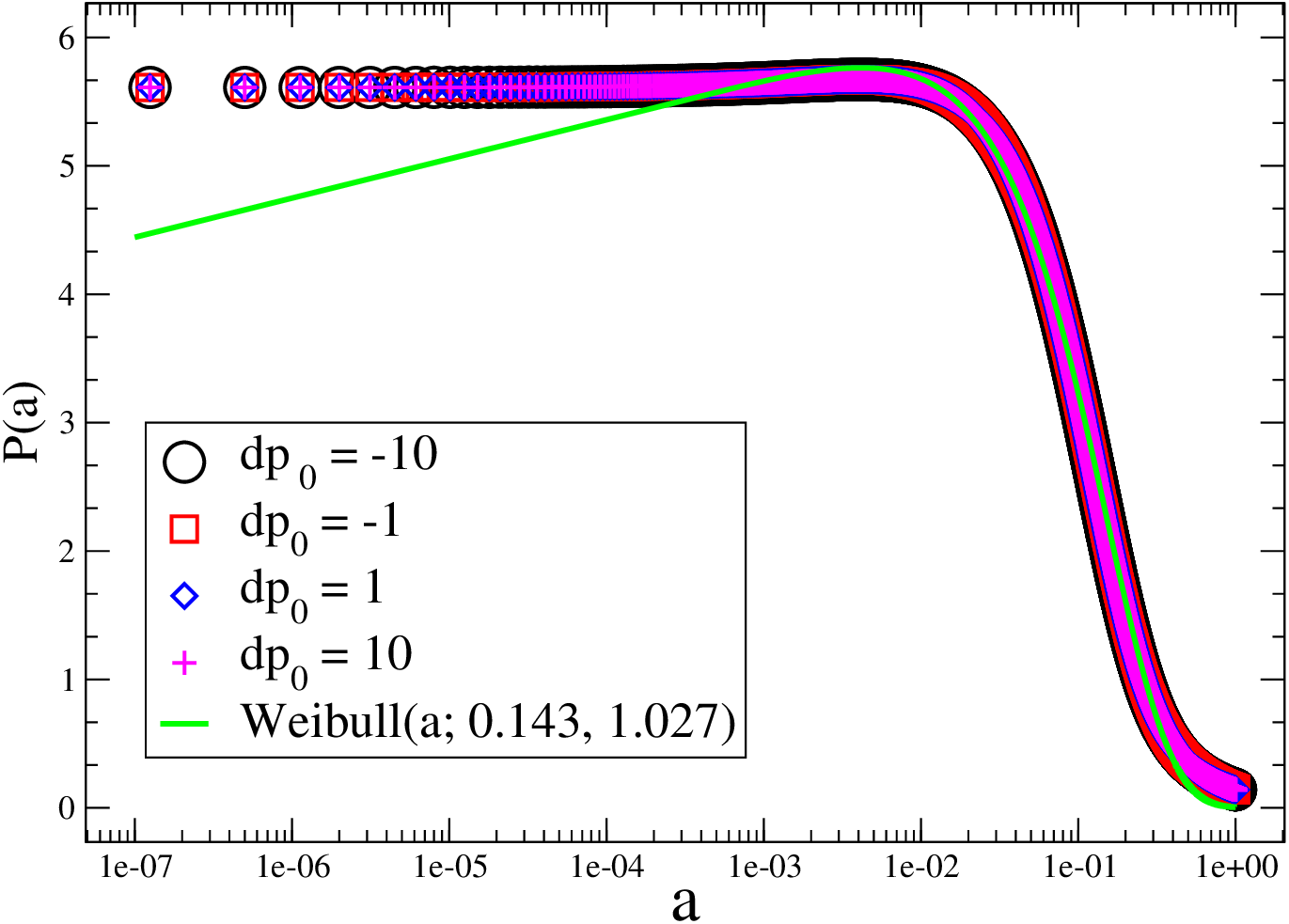}
    \caption{Degree distribution derived from the solution of the IVP \eqref{IVPY0}. The value of $\dd p_0$ does not interfere with the degree distribution. The result fits well with a Weibull distribution for a network with many \change{links}, $L > 10^4$.}
     \label{fig:IVPL2}
  \end{minipage}

\end{figure}

The degree distribution $P(a = \frac{k}{2L})$ obtained from this method is presented in Fig. \ref{fig:IVPLYMax}, where we see that, under normalization, the value of $\dd p_0$ does not influence the probability values. Furthermore, upon the rescalling $\bar{a} = a/a_{max}$ and $\bar{P} = P \sqrt{a_{max}}$, or similarly $\bar{y} = y/y_{max}$, the number of links $L$ does not alter the behaviour of the degree distribution, as seen in Fig. \ref{fig:IVPL32}.

Another initial value condition we investigated was to consider every node in the graph to have at least one \change{link},
\begin{equation}\label{IVPY0}
\pp(y=0) = 0 \qq{and} \eval{\dv{\pp}{y}}_{y=0} = \dd p_0 \ ,
\end{equation}
The integration runs until $y = \sqrt{8}$, where $a = 1$, and then normalized. Similarly to the previous case, the value of $\dd p_0$ does not interfere with the degree distribution after normalization, as shown in Fig. \ref{fig:IVPL2}.

\begin{table}%[h]
\centering
\begin{tabular}{|c|c|c|c|}
\hline
\change{Distribution} & $P(a) \propto $ & Range & RMSE\\ 
\hline \hline
Weibull($\bar{a}$; $\lambda$, $k$) & $\qty(\frac{\bar{a}}{k})^{\lambda-1} e^{-\qty(\frac{\bar{a}}{k})^{\lambda}}$ & $\bar{a}\in (0, 2\cdot 10^{-1}]$ in Fig. \ref{fig:IVPL32} & 0.045 \\
\hline

Gamma($\bar{a}$; $\lambda$, $k$) & $\bar{a}^{-\lambda} e^{-k \bar{a}}$ & $\bar{a} \in (0, 8\cdot 10^{-1}]$ in Fig. \ref{fig:IVPL32} & 0.166  \\
\hline

PLIEC($\bar{a}$; $\lambda$, $k$) & $\bar{a}^{-\lambda} e^{- k/\bar{a} }$ & $\bar{a} \in [6\cdot 10^{-2},\ 2\cdot 10^{-2}]$ in Fig. \ref{fig:IVPL32} & 0.155  \\
\hline

Power-Law($\bar{a}$; $\lambda$) & $\bar{a}^{-\lambda}$ & $\bar{a} \in (0, 5\cdot 10^{-3}]$ in Fig. \ref{fig:IVPL32} & 0.385  \\
\hline

Weibull($a$; $\lambda$, $k$) & $\qty(\frac{a}{k})^{\lambda-1} e^{-\qty(\frac{a}{k})^{\lambda}}$ & $a\in [10^{-4},10^0]$ in Fig. \ref{fig:IVPL2} & 0.118  \\
\hline
\end{tabular}

\caption{Summary of the degree distributions fit to the results of the IVPs \eqref{IVPYM} and \eqref{IVPY0}. The chosen range takes into account the value of the root mean square error (RMSE), which is minimized relative to the entire range of $a$ in the plot.}
\label{tab:results}

\end{table}

\change{
In Fig. \ref{fig:IVPL32} and Fig. \ref{fig:IVPL2} we fit our numerical results of $P(a)$ to some fat-tailed distributions. Namely we are interested in
(i) a regular power law obtained as a result of preferential attachment \cite{Barabasi99};
(ii) the power law with exponential cutoff -- Gamma distribution -- and (iii) the Weibull distribution, both reported to be found in real-world networks \cite{Broido19};
(iv) the power law with inverse exponential cutoff (PLIEC) found to be the result of an entropic coarse-graining on networks ensembles obtained from constraints on the node degrees \cite{Radicchi20}.} 
In Table \ref{tab:results} we present  the range -- region at which the distribution is valid, chosen based on the values of $a$ that minimize the root mean squared error (RMSE) between the functional form and the numerical results -- for which each distribution compares to our numerical result for $P(a)$.
%Inspired by the distributions reported to be found in real-world networks \cite{Broido19}, we fit the numerical results for $P(a)$ to the fat-tailed distributions in Table \ref{tab:results}. The range -- region at which the distribution is valid -- is chosen based on the values of $a$ that minimize the root mean squared error (RMSE) between the functional form and the numerical results. 
The fact that for Fig. \ref{fig:IVPL32} the solutions start from above zero means that the degree distribution is considered from the minimal possible non-zero connectivity, $k=1$. Also, the case where every node has at least one \change{link} only fits with a real-world degree distribution for networks with $L > 10^4$ links.
Note from \eqref{DegDistDef} that the metric leads to a natural power law behavior as $a^{-0.5}$, which was found throughout most of Fig.  \ref{fig:IVPL32} range.

\section{Conclusions}\paragraph{}
We presented an entropic dynamics of graphs with a fixed number of nodes $N$ and \change{links} $L$. This model leads to the Gibbs distribution \eqref{AiAj} whose information metric is given by \eqref{finalmetric}.
When the dynamical assumption is that we evolve the parameters continuously and constrained in the statistical manifold we are lead into the Fokker-Planck equation \eqref{FPeq}.
Steady-state solutions for two different IVPs are presented: The differential equation results for \eqref{IVPYM}  are graphically represented in Fig. \ref{fig:IVPLYMax} and fitted for Weibull and Gamma distributions in Fig. \ref{fig:IVPL32}. %\todoblue{Added this}, where there is a power law behavior due to the metric. 
Also, the results differential equation for \eqref{IVPY0} fitted for the Weibull distribution are presented in Fig.\ref{fig:IVPL2}, where the real-world distribution is only valid for graphs with many $L > 10^4$ links.

Our result is an information theory approach for dynamics of networks in which, under very general assumptions, the power law behavior emerges. Naturally, this work is not the single \change{dynamical process possible}. Under entropic dynamics, other random graph models can be studied and other constraints -- instead of or in addition to \eqref{eq:EDGConst} -- can be implemented when maximizing the dynamical entropy $\mathsf{S}$ as in \eqref{DynamicalS}.
This brings a perspective that the present article can be seen as an example on the ability to obtain dynamical process in complex systems using information theory. \change{Therefore, building on the work presented here, we believe entropic dynamics gives a framework able to address the  overarching need for dynamical models of networks.  }

\section*{Acknowledgments}\paragraph{}
We would like to thank G. Bianconi and  A. Caticha for insightful discussion in the development of this article.
We also thank B. Arderucio Costa, M. Rikard, and L. Sim\~oes  for their writing suggestions and proofreading. 

P.Pessoa was financed in part by  CNPq -- Conselho Nacional de Desenvolvimento Científico e Tecnológico-- (scholarship GDE 249934/2013-2)

%\newpage

%\bibliographystyle{unsrt}
\bibliographystyle{naturemag-doi}
\bibliography{referencias}

\end{document}